# The problem of choosing of the group of symmetry of paramagnetic phase in the theory of magnetic phase transitions and the exchange multiplets.

Khisa Sh. Borlakov[*] and Albert Kh. Borlakov

*North Caucasian State Humanitarian and Technological Academy,*

*36 Stavropolskaya str., Cherkessk, Russia, 369001*

It is shown that for the theoretical description of the magnetic and structural phase transitions in the magnetics with magnetoactive ions of transition metals within the framework of the Landau theory of phase transitions the exchange group GxO(3) should be chosen as the group of symmetry of paramagnetic phase. Such a choice allows to describe the transition from the paramagnetic phase into the isotropic magnetically ordered phase (at $T_c$), and the spin-orbit transition from the isotropic into the anisotropic phase (at $T < T_c$). The exchange multiplet describes the spin-orbit phase transition.

## 1. Introduction

The exchange multiplet is called a certain set of irreducible representations (IRs) of the space group of symmetry of a paramagnetic crystal (*G*) that is generated by the single IR of the exchange paramagnetic symmetry group *G*xO(3), where O(3) is the three-dimensional group of improper spin rotations. The concept of the exchange multiplet was introduced in [1] in order to use, at least indirectly, the concept of the soft mode in describing magnetic structures induced by several IRs of the space group of the paramagnetic crystal *G*. We will show that the concept of the exchange multiplet not only facilitates the formal description of magnetic structures but also has a deep physical meaning. Let's consider the physical meaning of the exchange multiplets, following [2].

## 2. Physical grounds for the choice of the symmetry group of a paramagnetic phase

Exchange multiplets only arise if we select the exchange paramagnetic group *G* x *0(3)* as the magnetic symmetry group of the paramagnetic phase rather-than the Shubnikov paramagnetic group *GV* (/' is the symmetry operation of spin inversion). It is assumed [1] that the Shubnikov paramagnetic group exactly describes the magnetic symmetry of the paramagnetic phase of any crystal, whereas the exchange paramagnetic group describes this symmetry only approximately; but this is not the case.

Let us consider this problem in more detail. The magnetism of solids is mainly caused by the presence of ions of *3d* elements, rare-earth 4f elements, or actinides (5f elements) in them [3]. These elements behave quite differently in crystals. The Hamiltonian of a *3d* ion can be written (in the order of decreasing contributions of the interactions) as follows [4]:

$$H = H_i + H_{cf} + H_{ls} \qquad (1)$$

The first summand corresponds to intra-atomic interactions; the second term is the interaction with the crystal field; and the third term is related to the spin-orbit interaction. Since the spin-orbit interaction is much weaker than the energy of the crystal field, there is no correlation between the direction of the spin and the electric charge density distribution (electron cloud) of the *3d* ion. In the paramagnetic phase, the spins have random directions independent of the directions of the crystal axes. Below the Curie temperature, a correlation appears between the spin directions caused by exchange interactions; i.e., a long-range magnetic order arises (magnetic ordering). If this is ferromagnetic ordering, there is no correlation between the direction of spontaneous magnetization and the crystallographic axes; the ferromagnetic phase is isotropic (absolutely magnetically soft). The symmetry of the paramagnetic phase must be described by the exchange paramagnetic group *G* x O(3).

For a rare-earth ion, the summands in the Hamiltonian are arranged in another order:

$$H = H_i + H_{ls} + H_{cf} \qquad (2)$$

The electron cloud of a *4f* ion, unlike a *3d* ion, is not spherical; it has a strongly anisotropic shape. When such an ion is introduced into a crystal lattice, the shape of the electron cloud remains virtually unaltered, since the spin-orbit interaction is much higher than the energy of the crystal field. The anisotropic electron cloud of a4f ion cannot be oriented in an arbitrary manner in a crystal field; the cloud rotates in such a manner that the energy of electrostatic interaction becomes minimum. Therefore, when a 4f ion is in a crystal lattice, there are directions of preferred orientation for its electron cloud and, consequently, a preferred orientation of the magnetic moment of the rare-earth ion, whose direction is rigidly correlated with the electron charge distribution. It does not follow from the above that the paramagnetic phase of a crystal containing rare-earth ions represents a "jellium" in which magnetic moments oriented along one of the preferred symmetry axes of the local surroundings are "frozen in." In reality, the situation is completely analogous to that characteristic of the dynamic Jahn-Teller (JT) effect in crystals. As is known [5, 6], the meaning of the dynamic JT effect consists in the fact that the ion state with an orbitally degenerate energy level is unstable and this degeneration is lifted-by the distortion of the coordination polyhedron in whose center the JT ion is situated. Jahn and Teller [7] believed that the meaning of the effect consisted in just these phenomena. However, as was shown by Bersuker and his school [6], the situation is somewhat different. If we consider a system consisting of a JT ion and a coordination polyhedron, we will see that there is no degeneration of the energy levels in such a system. The energy levels of the JT ion strongly interact (vibron interaction) with vibration energy levels of the coordination polyhedron, so that the orbital degeneration always proves to be lifted. This is what is called the dynamic JT effect. The asymmetric electron cloud of a 4f ion in a magnetic site of a crystal behaves similarly. At any time instant, the coordination polyhedron of a 4f ion proves to be distorted in such a manner as to minimize the electrostatic energy. There are several energetically equivalent types of distortion of the coordination polyhedron and corresponding directions of the orientation of the electron cloud. Between these states of the 4f ion + coordination polyhedron system, continuous transitions occur. The symmetry of the coordination polyhedron remains (on average in time) unaltered and is higher than the symmetry of the distorted polyhedron. This holds down to the Curie temperature. Below the Curie point, the term $E$ in the free energy $F = E - TS$ begins to dominate and the crystal transforms into a magnetically ordered state, which is accompanied by structural distortions of the crystal lattice. Unlike the *3d* ion, the 4f ion has a discrete set of various directions of orientations of the magnetic moment in space and these directions are related to the crystallographic axes even in the paramagnetic phase. Therefore, in order to obtain the magnetic symmetry group of the paramagnetic phase we may add only the operation of inversion of the direction of the magnetic moment (1') to the operations of the space group G, so that the symmetry group of the paramagnetic crystal with rare-earth ions is the Shubnikov group *G1'*.

Thus, the choice of the symmetry group of the paramagnetic phase depends on which ions are magnetically active: either the ions with *ls* coupling or ions with *jj* coupling; i.e., the choice of the symmetry group is in essence determined by dynamic rather than geometric factors. The choice of the symmetry group of the paramagnetic phase predetermines many physical consequences of the model of the magnetic state of the crystal.

### 3. The general method for calculating the exchange multiplet

Let the symmetry of the paramagnetic phase be characterized by the exchange group $G \times O(3)$. The order parameter that describes the transition into the magnetically ordered phase is transformed according to the IR $d_{\vec{k}}^{v} \times V'$ of this group [1], where $d_k$ is the IR of the space group G belonging to the star of the wave vector $\{\vec{k}\}$ and having an order number v, and $V$ is the vector IR of the three-dimensional group of spin rotations O(3). The IR $d_k$ is subject to the following restriction: it should enter into the so-called permutation representation of group $G$ at the sites of the crystal lattice occupied by magnetic atoms. The IR $d_{\vec{k}}^{v} \times V'$ is called the critical IR [8] and completely describes the magnetic symmetry of phases that arise below the Curie-Neel point. These phases do not possess magnetic anisotropy; i.e., they are isotropic. The anisotropic magnetic phases that arise as the temperature decreases further cannot be induced directly by the IR $d_{\vec{k}}^{v} \times V'$. In order to determine the critical IR that induces the transition from the isotropic to the anisotropic phase, the matrices of rotation by arbitrary angles in the IR $d_{\vec{k}}^{v} \times V'$ corresponding to the IR $V'$ of the group O(3) should be replaced by matrices that are consistent with the rotations of the space group. Actually, this will result in the replacement of the IR of the O(3) group by a pseudovector IR V of the group $G$ and the IR $d_{\vec{k}}^{v} \times V'$ will transform into the reducible representation $d^v{}_k$ x $V$ of the group $G$. It is this representation that forms the above multiplet, and the procedure we suggested is another method of calculating it. It is evident that the dimensionality of the multiplet is $3s$, where $s$ is the dimensionality of the IR $d_k$. The multiplet can be reduced using formulas given in [1], but we can proceed otherwise. Take the matrices of the IR $d_k$ and $V$ corresponding to the generators of the group $G$ (i.e., a few matrices of dimensionality 3s). Then we should find an orthogonal transformation that changes these matrices to a block-diagonal form. It is easier to do this for concrete IRs of space groups than to use the general group-theoretical formulas from [1]. Further investigation of exchange multiplets will more easily be conducted on a concrete example.

## 4. Isotropic antiferromagnetic phases in spinels whose magnetic unit cell coincides with the crystal-chemical unit cell

The symmetry of the crystal lattice of spinel is characterized by the space group $O_h^7 = Fd3m$. The number of atoms in the fcc unit cell is four times that in the primitive cell. Chemical compounds that crystallize in a spinel-type structure have the formula $AB_2O_4$, where $A$ and $B$ are metal cations and O are oxygen anions. The magnetic atoms occupy the tetrahedral sublattice (crystallographic positions 8(a)) and/or the octahedral sub-lattice (crystallographic positions *16(d)*); the anions occupy the crystallographic positions 32(*e*). In the figure, a fragment of the cubic unit cell of the spinel type is given, which is equivalent to the primitive unit cell. In this figure, the letters *A* and *B* denote two octants of the unit cell, whose staggered arrangement yields the Bravais cell. The vectors of the principal translations have the following coordinates *(a is the lattice parameter)*:

$$\vec{a}_1 = \frac{a}{2}\{1,1,0\}, \vec{a}_2 = \frac{a}{2}\{0,1,1\}, \vec{a}_3 = \frac{a}{2}\{1,0,1\}$$

The anions are shown in the figure by open circles; the atoms in octahedral positions, by solid circles; and tetrahedral atoms, by crossed circles. In order to calculate the scalar and pseudovector basis functions of the IR of group $O_h^7$, the atoms should be strictly numbered in sequence. We consider magnetic ordering in the octahedral sublattice; the order numbers of the octahedral atoms and their coordinates in fractions of the lattice parameter are as follows:

$$16(d): 1(\frac{5}{8}\frac{1}{8}\frac{1}{8}), 2(\frac{5}{8}\frac{3}{8}\frac{3}{8}), 3(\frac{7}{8}\frac{1}{8}\frac{3}{8}), 4(\frac{7}{8}\frac{3}{8}\frac{1}{8})$$

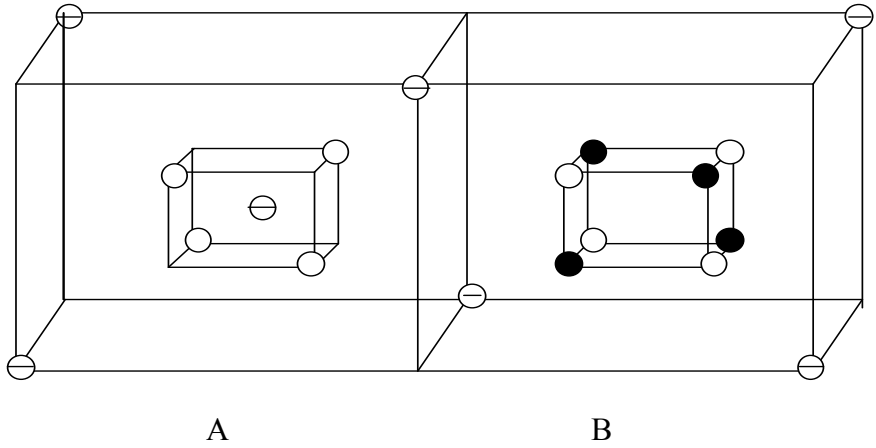

A  B

**Figure.**

Let us consider magnetic structures whose unit cell coincides with the crystal-chemical cell. All these structures are described by IRs belonging to the star of the wave vector $\vec{k}_{11} = 0$. The IR of the space group belonging to the zero wave vector and the IR of the corresponding point group are isomorphic; therefore, we will use the "chemical" symbols of the IR of the point group $O_h$ in what follows [9]. Of ten IRs, that enter into the star of this wave vector, only the following three 11-1($A_{1g}$), 11-4($A_{2u}$), and 11-7($F_{2g}$) - enter into the permutation representation at the sites 8($a$) and 16($d$). Two one-dimensional IRs $A_{1g}$ and $A_{2u}$, generate exchange singlets $A_{1g} \times \widetilde{V} = F_{1g}$ and $A_{2u} \times \widetilde{V} = F_{2u}$. The transition from the paramagnetic to therromagnetic phase occurs through the identical IR $A_{1g}$. As the temperature decreases further, the transition from the isotropic to the anisotropic ferromagnetic phase occurs through the IR $A_{1g} \times \widetilde{V} = F_{1g}$. The transition to the collinear isotropic antiferromagnetic phase occurs through the IR $A_{2u} \times V'$. A further decrease in the temperature causes the transition to the anisotropic antiferromagnetic phase through the IR $A_{2u} \times \tilde{V} = F_{2u}$. The antiferromagnetic structure of this phase is caused by the 1 : 1 ordering of spins in the tetrahedral sublattice.

Finally, we turn to the most interesting case - the magnetic ordering induced by the critical IR $F_{2g} \times V'$ of the exchange paramagnetic group $O_h$ x O(3). This IR induces a complex three-dimensional anti-ferromagnetic ordering on octahedral sites. The vectors of the magnetic moment on the sites can be calculated by the formula

$$\vec{S}(\vec{r}) = \sum_{a=1}^{3} \sum_{i=1}^{n} S_a^i \varphi_i(\vec{r}) \cdot \vec{e}^a = \sum_i \vec{S}_i \varphi_i(\vec{r}), \qquad (3)$$

where $\varphi_i(\vec{r})$ is the scalar basis function of the IR $F_{2g}$, $\vec{e}^a$ is the orthonormalized basis in the three-dimensional spin space, $S_a^i$ is the order parameter referring to the *a*th component of the spin vector $\vec{S}(\vec{r})$, and $\vec{S}_i$ is the column vector consisting of $S_a^i$. As is known [8], the conventional n-component order parameter referring to a certain n-dimensional IR of the space group is called the stationary vector. The magnetic ordering in the isotropic phase is described, as is seen from (3), by three sets of conventional stationary vectors, i.e., by a matrix of stationary vectors of dimensionality $3 \times n$. The isotropic antiferromagnetic phases induced by the critical IR $F_{2g} \times V'$ are given in Table 1.

Table 1. Isotropic antiferromagnetic phases induced by the critical irreducible representation (IR) $F_{2g} \times V'$ of group $O_h^7 \times O(3)$

| $\vec{S},\vec{S},\vec{S}$ | $\vec{S},0,0$ | $\vec{S}_1,\vec{S}_2,\vec{S}_2$ | $\vec{S}_1,\vec{S}_2,\vec{S}_3$ |
|---|---|---|---|
| $D_{3d}^5$ | $D_{2h}^{28}$ | $C_{2h}^2$ | $C_i^1$ |

For calculations using (3), we also need scalar basis functions for the *16(d)* positions; these are listed in Table 2. This table also contains the results of calculations of the spin moment at each of the atoms for the isotropic magnetic phases given in Table 1.

It can be seen from Table 2 that the isotropic phase (S, 0, 0) is characterized by a 1 : 1 spin ordering on the octahedral sublattice and the (S, S, S) phase, by a 1 : 3 spin ordering (similar to the types of atomic ordering [9]). Thus, we considered isotropic phases generated by three IRs, namely, 11-1 *($A_{1x}$)*, 11-4 (A₂,,), and 11-7 *($F_{2x}$)*, belonging to the star of the zero wave vector. The "true" exchange multiplet is only generated by the 11-7 IR, so we will consider only this IR in what follows.

Table 2. Scalar basis functions of the IR $F_{2g}$ and the values of spin moments for magnetic atoms in crystallographic positions 16(d)

| Order numbers of the atoms → | 1 | 2 | 3 | 4 |
|---|---|---|---|---|
| $\varphi_1$ | 1 | -1 | -1 | 1 |
| $\varphi_2$ | 1 | -1 | 1 | -1 |
| $\varphi_3$ | 1 | 1 | -1 | -1 |
| $\vec{S},\vec{S},\vec{S}$ | $3\vec{S}$ | $-\vec{S}$ | $-\vec{S}$ | $-\vec{S}$ |
| $\vec{S},0,0$ | $\vec{S}$ | $-\vec{S}$ | $-\vec{S}$ | $\vec{S}$ |
| $\vec{S}_1,\vec{S}_2,\vec{S}_2$ | $\vec{S}_1 + 2\vec{S}_2$ | $-\vec{S}_1$ | $-\vec{S}_1$ | $\vec{S}_1 - 2\vec{S}_2$ |
| $\vec{S}_1,\vec{S}_2,\vec{S}_3$ | $\vec{S}_1 + \vec{S}_2 + \vec{S}_3$ | $-\vec{S}_1 - \vec{S}_2 + \vec{S}_3$ | $-\vec{S}_1 + \vec{S}_2 - \vec{S}_3$ | $\vec{S}_1 - \vec{S}_2 - \vec{S}_3$ |

## 5. Anisotropic antiferromagnetic phases in spinels induced by critical irreducible representation $F_{2g} \times V'$ of the $O_h^7 \times O(3)$ group

As was said above, the critical IR $F_{2g} \times V'$ itself only induces isotropic antiferromagnetic phases. In order to study the anisotropic phases, we must restrict this IR to the space group $O_h^7$; i.e., we should consider the exchange multiplet generated by this IR. Calculations using the general group-theoretical formula given in [1] yield the following result:

$$[F_{2g} \times V'] = F_{2g} \times \widetilde{V} = A_{2g} + E_g + F_{1g} + F_{2g}. \qquad (4)$$

Note that we designated the critical IR and the corresponding multiplet differently. As was expected, the 9-dimensional reducible representation decomposed into the direct sum of two 3-dimensional, one 2-dimensional, and one 1-dimensional IRs of the $O_h^7$ group. All IRs entering into the right-hand side of formula (4) are even with respect to spatial inversion and therefore can describe magnetic ordering.

Table 3. Complete condensation of stationary vectors of the critical IR $F_{1g}$

| $F_{1g}$ | $G_D$ | $A_{1g}$ | $A_{2g}$ | $E_g$ | $F_{2g}$ |
|---|---|---|---|---|---|
| ccc | $C_{3i}^2$ | a | a | | aaa |
| 0cc | $C_{2h}^3$ | a | | 0a | abb |
| c00 | $C_{4h}^6$ | a | | a,-$\sqrt{3}$ a | |
| $c_1 c_2 c_3$ | $C_i^1$ | a | a | ab | abc |

The set of IRs that describe a phase transition forms the so-called condensation [7]. One of these IRs determines the decrease in the magnetic or crystal symmetry and is called the critical IR; all the others describe secondary (related) phenomena compatible with the symmetry of the newly arising phases. Note that the exchange multiplet (4) coincides with the complete condensation generated by the critical IR $F_{1g}$, and with the complete condensation generated by the critical IR $F_{2g}$. Omitting the calculation details, we only give the final results and make comments of a physical nature.

Table 4. Complete condensation of stationary vectors of the critical IR $F_{2g}$

| $F_{2g}$ | $G_D$ | $A_{1g}$ | $A_{2g}$ | $E_g$ | $F_{1g}$ |
|---|---|---|---|---|---|
| $ccc$ | $D_{3d}^5$ | a | | | |
| $c_1 c_2 c_2$ | $C_{2h}^3$ | a | | a, $-\sqrt{3}$ a | $0a\bar{a}$ |
| $c00$ | $D_{2h}^{28}$ | a | | a, $-\sqrt{3}$ a | |
| $c_1 c_2 c_3$ | $C_i^1$ | a | a | ab | abc |

Table 3 gives the complete condensation of the critical IR $F_{1g}$, and Table 4 contains the complete condensation of the critical IR $F_{2g}$. Note that we designated the critical order parameter by a different letter and, unlike Table 1, its components are no longer vectors, which has a certain physical meaning. The complete condensation also contains the identity IR $A_{1g}$, which was absent in the exchange multiplet. Thus, a question that arises is which of the two IRs entering into the exchange multiplet - $F_{1g}$ or $F_{2g}$ - is critical. Unfortunately, we have no formal logical answer to this question and we resort to physical arguments. These arguments are as follows. Between the isotropic magnetic phases that are induced by the IR $F_{2g} \times V'$ and the anisotropic phases induced by one of the above IRs, a second-order phase transition should occur. For this to be possible, it is necessary that these phases have the same structure type and differ in the presence or absence of magnetic anisotropy. To facilitate comparison with Table 2, let us calculate the magnetic moments of the octahedral atoms in the anisotropic phases. The needed pseudovector basis functions for the 16(*d*) positions are given in Table 5. It can easily be verified, using stationary vectors from Tables 3 and 4, that the succession of magnetic structures in the isotropic and anisotropic phases is fulfilled only if the critical IR is $F_{1g}$.

Table 5. Pseudovector basis functions of crystallographic positions 16(J) for the IRs $F_{1\sigma}$ and $F_{2\sigma}$ [I]

| Order numbers of the atoms → ↓ IRs | 1 | 2 | 3 | 4 |
|---|---|---|---|---|
| $F_{2g}$ | 0 1 -1<br>-1 0 1<br>1 -1 0 | 0 -1 1<br>1 0 1<br>-1 -1 0 | 0 -1 -1<br>1 0 -1<br>1 1 0 | 0 1 1<br>-1 0 -1<br>-1 1 0 |
| $F_{1g}$ | 100<br>010<br>001 | 100<br>010<br>001 | 100<br>010<br>001 | 100<br>010<br>001 |
| $F'_{1g}$ | 011<br>101<br>110 | 0 -1 -1<br>-1 0 1<br>-1 1 0 | 0 -1 1<br>-1 0 -1<br>1 -1 0 | 0 1 -1<br>1 0 -1<br>-1 -1 0 |

As can be seen from Table 5, the critical IR $F_{1g}$ enters into the pseudovector representation on the octahedral sites doubly (two independent sets of basis functions correspond to it). The first set of basis functions corresponds to magnetic ordering in both the tetrahedral and octahedral sites. This critical IR originates from the critical IR $A_{1g} \times V'$ of the exchange paramagnetic group $O_h^7 \times O(3)$. Restricting this IR to the $O_h^7$ group, we obtain the exchange singlet $F_{1g}$. In this case, both the isotropic and the anisotropic phases turn out to be ordered ferromag-netically. The group theory was so provident as to prepare another set of pseudovector basis functions, which we designated $F'_{1g}$. It is these functions that correspond to the complex antiferromagnetic ordering on octahedral sites, and this ordering originates from the antiferromagnetic ordering induced by the critical IR $F_{2g} \times V'$ of the $O_h^7 \times O(3)$ group. The magnetic moments at octahedral atoms calculated by the formula

$$\vec{m}(\vec{r}) = \sum_i c_i \cdot \vec{\varphi}_i(\vec{r})$$

are given in Table 6. Similar calculations using basis functions of the critical IR $F_{2g}$ yield magnetic moments of atoms given in Table 7. It can be seen that the magnetic moments of atoms in the trigonal anisotropic phase are zero, whereas in the isotropic phase a 1 : 3 ordering should be observed (line S, S, S in Table 2). In Table 6, the trigonal phase corresponds to 1 : 3 ordering, as in the corresponding isotropic phase. In the tetragonal phase (c, 0, 0), the spins of atoms 1 -2 and 3-4 are antiparallel pairwise, similar to the corresponding isotropic phase (S, 0, 0). However, the isotropic phase was collinear antiferromagnetic, whereas the tetragonal (c, 0, 0) phase from Table 6 is noncollinear; nevertheless, both phases reveal a 1 : 1 spin ordering. The symmetry groups $G_D$ shown in Table 3 are the space groups of the corresponding anisotropic magnetic phases. Note that the choice of $F_{1g}$ as the critical IR is supported by the fact that phases that correspond to it have a lower symmetry than the phases corresponding to the critical IR $F_{2g}$.

Table 6. Magnetic moments of the octahedral atoms in the primitive unit cell of spinel for the critical IR $F_{1g}$

|     | 1 | 2 | 3 | 4 |
|-----|---|---|---|---|
| ccc | 2c(1,1,1) | 2c(-1,0,0) | 2c(0,-1,0) | 2c(0,0,-1) |
| 0cc | c(2,1,1) | c(-2,1,1) | c(0,-1,-1) | c(0,-1,-1) |
| c00 | c(0,1,1) | c(0,-1,-1) | c(0,-1,1) | c(0,1,-1) |
| abc | (b+c,a+c,a+b) | (-b-c-,a+c,-a+b) | (-b+c,-a-c,a-b) | (b-c,a-c,-a-) |

Table 7. Magnetic moments of the octahedral atoms in the primitive unit cell of spinel for the critical IR $F_{2g}$

|     | 1 | 2 | 3 | 4 |
|-----|---|---|---|---|
| ccc | 0 | 0 | 0 | 0 |
| acc | (0,a-c,-a+c) | (0,-a-c,a+c) | (2c,-a+c,-a-c) | (-2c,a+c,a-c) |
| c00 | c(01,-1) | c(0,-1,1) | c(0,-1,-1) | c(0,1,1) |
| abc | (-b+c,a-c,-a+b) | (b-c,-a-c,a+b) | (b+c,-a+c,-a-b) | (-b-c,a+c,a-b) |

Since the ferromagnetic ordering also corresponds to the same critical IR $F_{1g}$, we believe that the symmetry of the crystal lattices of the antiferromagnetic phases cannot be higher than the symmetry of the lattices of the ferromagnetic phases. If we had chosen the critical IR $F_{2g}$, it would have turned out that the complex antiferromagnetic ordered arrangements in the anisotropic phases were characterized by a higher symmetry of the crystal lattice than in the ferromagnetic phase. For example, the trigonal ferromagnetic ordering is characterized by the symmetry group $C_{3i}^2$, and the corresponding trigonal antiferromagnetic ordering that occurs by the IR $F_{2g}$ is characterized by the group $D_{3d}^5$.

Thus, the Landau postulate "that the phase transition occurs through one critical IR is also confirmed for magnetic phase transitions in exchange magnets. The exchange multiplet that is obtained from the critical magnetic IR of the exchange paramagnetic group always contains an IR that is an indisputable candidate for the critical IR.

**6. Conclusion**

The results of this work suggest the following. The concept of exchange multiplets, which was introduced in [1] as a convenient technique for considering magnetic structures described by several IRs of the space group, proves to have a deep physical meaning. All magnets are divided into two wide classes: exchange and nonexchange. In the first class, the magnetically active atoms are elements of the iron group, whose magnetic shell *(d* shell) is characterized by *Is* coupling. In these magnets, the transition to a magnetically ordered phase and the appearance of anisotropic magnetic properties occur in two stages at two different temperatures. In these materials, an isotropic magnetic phase exists. It is for describing just such magnets that the introduction of exchange multiplets is needed. The nonexchange magnets behave classically, i.e., as is described in textbooks on magnetism. The magnetic atoms in them are 4/ or 5/ elements, whose magnetic shells reveal *jj* coupling. In these magnets the transition to a magnetically ordered phase is accompanied by the simultaneous appearance of magnetic anisotropy. In order to describe the symmetry of magnetically ordered phases of nonexchange magnets, there is no necessity of introducing exchange multiplets. The appearance of exchange multiplets is a consequence of the fact that the symmetry of the paramagnetic phase is objectively described by the exchange paramagnetic group $G \times 0(3)$. The symmetry of the paramagnetic phase of nonexchange magnets is described by Shubnikov paramagnetic groups $GI'$. There is no doubt that the Landau postulate concerning the phase transition through the one critical IR also holds for magnetic transitions. Indeed, the exchange multiplet enters into the complete condensation of a certain critical IR. If there are two or more equivalent candidates for the role of the critical IR of those that enter into the exchange multiplet, then, as a rule, there are additional symmetry-related arguments in favor of the unambiguous choice of one of these as the critical IR. Thus, the magnets also admit the existence of a soft mode whose condensation causes a phase transition at the Curie point.

———————————————

*Electronic address: borlakov@mail.ru

1. Izyumov, Yu.A., Naish, V.E., and Ozerov, R.P., Neutron Diffraction of Magnetic Materials; Consultants Bureau: New York, NY, USA, 1991.

2. Borlakov, Kh.Sh., On the Physical Meaning of Exchange Multiplets, *Phys. Met. Metallogr.*. 1998, vol. 86, no. 2, pp. 123-128.

3. White, R., *Quantum Theory of Magnetism,* Heidelberg: Springer, 1983.

4. Krupicka, S., *Physik der Ferrite und der verwandten magnetischen Oxide* (Physics of Ferrites and Related Magnetic Oxides), Prague: Academia, 1973.

5. Bersuker, I.B., The Jahn-Teller Effect and Vibronic Interactions in Modern Chemistry, N.-Y.: Plenum press, 1984.

6. Jahn, G.A. and Teller, E., Stability of Multiatomic Molecules with Degenerate Electron States: I. Orbital Degeneration, *Proc. R. Soc. London A,* 1937, vol. 161, p. 220.

7. Sakhnenko, V.P, Talanov, V.M., and Chechin, G.M., Group-Theoretic Analysis of the Full Condensed Complex Arising upon Structural Phase Transformations, *Fiz. Met. Metalloved.,* 1986, vol. 62, pp. 847-856.

8. Landau, L.D. and Lifshitz, E.M., Quantum Mechanics: Non-Relativistic Theory 3rd edn (London: Pergamon) 1977

9. Talanov, V.M. and Chechin, G.M., *Teoretiko-gruppovoi raschet uporyadocheniya faz kristallov so strukturoi shpineli* (Group-Theoretic Calculation of Ordering of Crystals with the Spinel Structure), Available from VINITI, 1987, Moscow, no. 6517-V87.